\documentclass{svjour3}                     
\smartqed  
\usepackage{graphicx}
\usepackage{latexsym}
\usepackage{marvosym}
\usepackage{times}
\usepackage{graphicx}
\usepackage{subcaption}
\usepackage[compatibility=false]{caption}
\usepackage{mathrsfs}
\usepackage{amsmath}
\usepackage{amssymb}

\usepackage{tikz}
\usepackage{listings} 
\PassOptionsToPackage{rgb}{xcolor}
\usepackage{pgfplots}

\begin{document}

\title{Adjoint-based Optimal Flow Control for Compressible DNS
}
\subtitle{Open Cavity Noise Control}


\author{J. Javier Otero \and Ati S. Sharma \and Richard D. Sandberg}

\authorrunning{J.J. Otero, A.S. Sharma and R.D. Sandberg} 

\institute{J. Javier Otero (\Letter) \and Ati S. Sharma \at Faculty of Engineering and the Environment, University of Southampton, Southampton SO17 1BJ, U.K. \\
              \email{j.j.otero-perez@soton.ac.uk}            \\
           \and
           Richard D. Sandberg \at
              Department of Mechanical Engineering, University of Melbourne, Victoria 3010, Australia
}

\date{March 18, 2016}

\maketitle

\begin{abstract}
A novel adjoint-based framework oriented to optimal flow control in compressible direct numerical simulations is presented. Also, a new formulation of the adjoint characteristic boundary conditions is introduced, which enhances the stability of the adjoint simulations. The flow configuration chosen as a case study consists of a two dimensional open cavity flow with aspect ratio $L/H=3$ and Reynolds number $Re=5000$. This flow configuration is of particular interest, as the turbulent and chaotic nature of separated flows pushes the adjoint approach to its limit. The target of the flow actuation, defined as cost, is the reduction of the pressure fluctuations at the sensor location. To exploit the advantages of the adjoint method, a large number of control parameters is used. The control consists of an actuating sub-domain where a two-dimensional body force is applied at every point within the sub-volume. This results in a total of $2.256 \cdot 10^6$ control parameters. The final actuation achieved a successful reduction of the cost of $79.6\%$, by altering the directivity of the sound radiated by the trailing edge of the cavity and breaking up the large shear layer instabilities into smaller structures.
\keywords{Adjoint \and Optimal flow control \and Compressible DNS \and Cavity noise}
\end{abstract}

\section{Introduction}
\label{intro}
The control of separated flows has caught the attention of numerous researchers for several decades. The current computational capabilities permit the use of more advanced techniques for their study, which give a new perspective to develop flow control schemes. The use of the adjoint method to fluid flows is gaining in popularity amongst the flow control community to carry out gradient-based optimal flow control. Examples of successful applications of adjoint-based optimal flow control are turbulence reduction in a channel flow \cite{bewley}, aero-acoustic jet noise reduction \cite{bewley2,bodony2}, or optimal porous media distribution in an aerofoil's trailing edge to minimise aero-acoustic radiation \cite{schulze2}. Note that due to the high computational demands of the method, it is a common practice to make simplifications of the flow, such as assuming incompressibility \cite{adj_inc_BFS}, constant viscosity \cite{bodony} or even linearising the dynamics of the flow \cite{schmid_mfree}. \\

\begin{figure}
	\centering
	\begin{subfigure}[b]{0.49\textwidth}
		\centering
		\subcaption{}
	    \definecolor{c808080}{RGB}{128,128,128}
\definecolor{cffccaa}{RGB}{255,204,170}
\definecolor{cb3b3b3}{RGB}{179,179,179}

\begin{tikzpicture}[trim axis left, trim axis right]
			\begin{axis}[
			enlargelimits=false,axis on top,
			scale only axis,
			xlabel={$x$},
			ylabel={$y$},
			every axis y label/.style={at={(current axis.west)},rotate=90,above=0.40cm},
			width=6.6cm,
			height=2.75cm,
			xmin=-1.5,xmax=4.5,
			ymin=0,ymax=2.5
			]
			\addplot graphics [xmin=-1.5,xmax=4.5,ymin=0,ymax=2.5]		{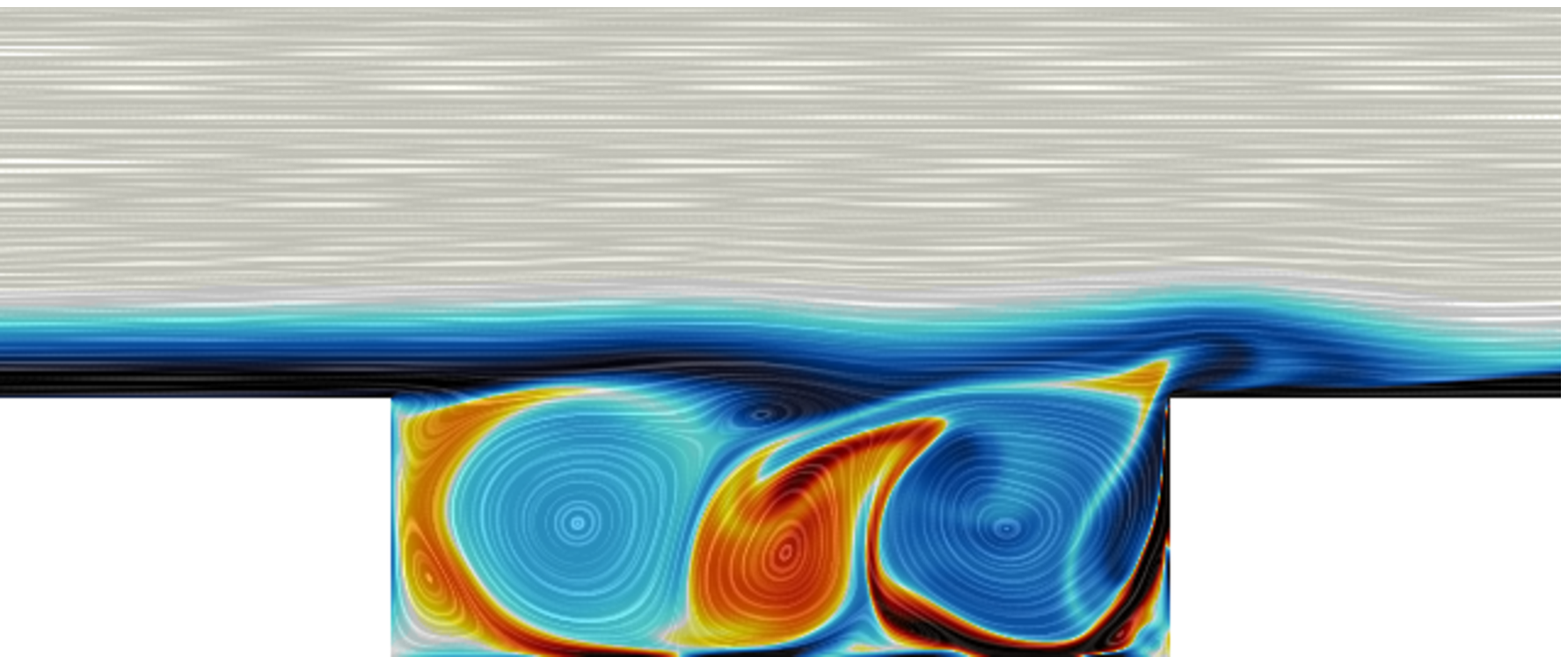};
			
    		 
			\addplot[color=red!80,
			 densely dotted,
            samples=51,
            domain=0:-3*pi/2,
            line width=0.75pt]
            ({3+0.25*cos(deg(x))}, 
             {1-0.25*sin(deg(x))});
             
  			\addplot[color=red!80,
			 densely dotted,
            samples=51,
            domain=0:-3*pi/2,
            line width=0.75pt]
            ({3+1*cos(deg(x))}, 
             {1-1*sin(deg(x))});         
             
  			\addplot[color=red!80,
			 densely dotted,
            samples=51,
            domain=0:-3*pi/2,
            line width=0.75pt]
            ({3+1.750*cos(deg(x))}, 
             {1-1.750*sin(deg(x))});      
             
  			\addplot[color=red!80,
			 densely dotted,
            samples=51,
            domain=0:-3*pi/2,
            line width=0.75pt]
            ({3+2.50*cos(deg(x))}, 
             {1-2.50*sin(deg(x))});        
             
  			\addplot[color=red!80,
			 densely dotted,
            samples=51,
            domain=0:-3*pi/2,
            line width=0.75pt]
            ({3+3.250*cos(deg(x))}, 
             {1-3.250*sin(deg(x))});       
             
  			\addplot[color=red!80,
			 densely dotted,
            samples=51,
            domain=0:-3*pi/2,
            line width=0.75pt]
            ({3+4*cos(deg(x))}, 
             {1-4*sin(deg(x))});                                       

			 \addplot[patch,patch type=rectangle,c808080!50!black,shader=flat]
    		 coordinates { (-1.5,0) (0,0) (0,1) (-1.5,1)};
    		 \addplot[patch,patch type=rectangle,c808080!50!black,shader=flat]
    		 coordinates { (3,0) (4.5,0) (4.5,1) (3,1)};    
    
    \end{axis}
\end{tikzpicture}
	    \label{fig:flow_setup_a}
	\end{subfigure}
	~
	\centering
	\begin{subfigure}[b]{0.49\textwidth}
		\subcaption{}
		\centering
	    \definecolor{c808080}{RGB}{128,128,128}
\definecolor{cffccaa}{RGB}{255,204,170}
\definecolor{cb3b3b3}{RGB}{179,179,179}

\begin{tikzpicture}[trim axis left, trim axis right]
			\begin{axis}[
			enlargelimits=false,axis on top,
			scale only axis,
			xlabel={$x$},
			every axis y label/.style={at={(current axis.west)},rotate=90,above=0.9cm},
			width=6.6cm,
			height=2.75cm,
			xmin=-1.5,xmax=4.5,
			ymin=0,ymax=2.5
			]
			\addplot graphics [xmin=-1.5,xmax=4.5,ymin=0,ymax=2.5]		{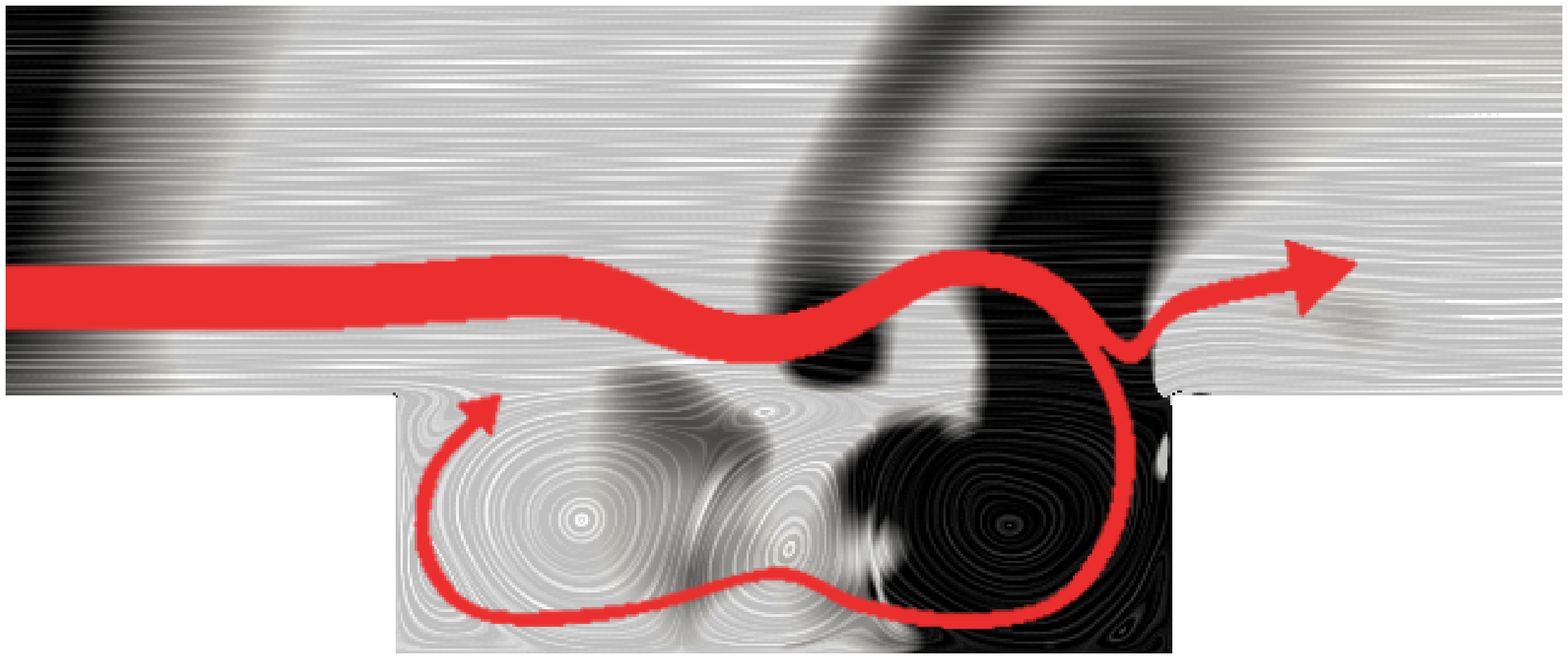};
             
			\addplot[color=blue!90,
			 densely dotted,
            samples=51,
            domain=0:2*pi,
            line width=1pt]
            ({4.1+0.25*cos(deg(x))}, 
             {1.6-0.25*sin(deg(x))});   

             \addplot[patch,patch type=rectangle,green!60,opacity=0.7,shader=flat]
    		 coordinates { (-0.5,1.20) (-0.25,1.20) (-0.25,1.54) (-0.5,1.54)};          
             
			 \addplot[patch,patch type=rectangle,c808080!50!black,shader=flat]
    		 coordinates { (-1.5,0) (0,0) (0,1) (-1.5,1)};
    		 \addplot[patch,patch type=rectangle,c808080!50!black,shader=flat]
    		 coordinates { (3,0) (4.5,0) (4.5,1) (3,1)};    
    
    \end{axis}
\end{tikzpicture}
	    \label{fig:flow_setup_b}
	\end{subfigure}
	\caption{\textbf{(a)} Instantaneous contours of $z$-vorticity. Red dashed lines represent the sound radiation from the trailing edge. \textbf{(b)} Sketch of the flow control setup with an instantaneous snapshot of the flow dilatation field in the background. The green box and the blue dashed circle depict the flow actuation and sensing regions, respectively.}
\label{fig:flow_setup}
\vspace{-0.5cm}
\end{figure}

\noindent In the current investigation we present a recently developed multi-block full compressible Navier-Stokes continuous adjoint solver with non-constant viscosity. The code is applied to a two dimensional separated flow scenario to explore the capabilities and limitations of the method. In particular, we focus on the importance of the sensor and actuator locations. Additionally, the derivation of the method and especially a novel formulation of the adjoint characteristic boundary conditions are explained in detail, which adds to this article a pedagogical value. The flow configuration chosen for the present investigation is an open cavity flow with aspect ratio $L/H = 3$ (figure \ref{fig:flow_setup_a}). Despite its simple geometry, this flow configuration gives rise to complex non-linear flow phenomena with constant flow separation and constant reattachment. Kelvin-Helmholtz type instabilities grow in the separated shear layer, creating non-linear structures which impinge on the trailing edge of the cavity. This unsteady interaction radiates acoustic waves that also propagate upstream. Due to the high receptivity of the leading edge, these compressible flow events further excite some of the unstable shear layer modes. The concatenation of these phenomena results in a feedback loop that is responsible for self-sustained oscillations, which are also known as Rossiter's modes \cite{rossiter_modes}. For further reading about open cavity flows, a summary of the recent computations and experimental techniques can be found in \cite{Lawson2011186}.\\

\noindent In summary, the present investigation uses adjoint-based optimal flow control on a cavity flow to minimise the acoustic radiation at the sensor location actuating on the upstream flow (figure \ref{fig:flow_setup_b}). The control methodology is introduced in section \ref{sec:flow_cont_opt}, where the different branches within adjoint-based optimisation and their limitations are discussed. Sections \ref{sec:3} and \ref{sec:4} describe the governing equations and the numerical treatment employed for the flow and adjoint simulations respectively. The flow control exercise alongside the results is described in section \ref{sec:5}. Finally, section \ref{sec:6} presents the conclusions.


\section{Flow Control and Optimization}
\label{sec:flow_cont_opt}

To evaluate the performance of a flow control strategy a generic quadratic cost functional is defined as \begin{equation}
\mathcal{J}\left(Q,\phi\right) = \frac{1}{2} \int_0^T \int_{\Omega} \left[ W\left(\vec{x}\right) Q^{\text{T}} \mathbf{M}_1 Q + \phi^{\text{T}} \mathbf{M}_2 \phi \right] \mathrm{d}\vec{x}\mathrm{d}t,
\end{equation} where $Q$ represents the state variables and $\phi$ is the vector of control parameters. $\mathbf{M}_1$ sets the relation of the state variables in the cost function and $\mathbf{M}_2$ accounts for the cost in the actuation parameters. The function $W\left(\vec{x}\right)$ weights the cost in the domain $\Omega$. The optimal control will be the values of the actuation parameters that result in the absolute minimum of the cost functional for the time horizon $T$.\\

\noindent To drive the cost function towards its minimum, the gradient \begin{equation}
\frac{D\mathcal{J}}{D\phi} = \frac{\partial \mathcal{J}}{\partial Q} \frac{\mathrm{d}Q}{\mathrm{d}\phi} + \frac{\partial \mathcal{J}}{\partial \phi}
\label{eq:grad_0}
\end{equation} must be computed. The challenging part of obtaining the gradient resides in the computation of the \textit{sensitivities} of the system $\mathrm{d}Q / \mathrm{d}\phi$. The most intuitive method to calculate this term is perturbing individually each of the control parameters, and then using a finite difference scheme to approximate the derivatives. This process would be repeated until the minimum value of the cost functional is reached, where $D\mathcal{J}/D\phi = 0$. With this method, the amount of required fluid simulations increases proportionally with the number of control parameters, which makes this approach unattainable when dealing with practical CFD simulations \cite{schulze1}. Fortunately for the present investigation, there are alternatives which provide optimal control with a much lower computational cost. \\

\subsection{Adjoint Method}
\label{subsec:adj_method}

\noindent The Lagrangian functional $\mathcal{L}$ is defined as \begin{equation}
\mathcal{L} \left(Q,Q^*,\phi \right) = \mathcal{J} \left( Q, \phi \right) - \left({Q^*}\right)^\text{T} \cdot \mathcal{N}\left(Q,\phi \right),
\label{eq:LFunc}
\end{equation} where $\mathcal{N}$ is the operator representing the set of governing equations and $Q^*$ are the \textit{co-state} or \textit{adjoint} variables. The governing equations, the adjoint equations and the optimality condition can be derived from the Lagrangian functional \cite{gunz} by setting its first variation with respect to $Q^*$, $Q$ and $\phi$ equal to zero, respectively
\begin{align}
\frac{\delta \mathcal{L}}{\delta Q^*} &= 0 \qquad \Rightarrow \qquad \mathrm{State \  equations}\\
\label{eq:def_adj_eq}
\frac{\delta \mathcal{L}}{\delta Q} &=0 \qquad \Rightarrow \qquad \mathrm{Adjoint \ equations}\\
\frac{\delta \mathcal{L}}{\delta \phi} &=0 \qquad \Rightarrow \qquad \mathrm{Optimality \ condition}.
\end{align}

\noindent From these three systems of equations the optimal control can be achieved by solving iteratively the state equations, followed by the solution of the adjoint equations, to finally use these two results to compute the updated control parameters with the optimality condition. Unfortunately, this method is equivalent to a steepest descent algorithm with a fixed step size \cite{gunz2}, which can converge very slowly. Due to the great cost of performing DNS, we instead use the L-BFGS algorithm \cite{lbfgsb}, which is more efficient \cite{Badreddine20141}. Note that the tolerances of the optimisation algorithm might require modification to satisfy the Wolfe conditions \cite{bfgs}. The authors recommend a low tolerance in the descent condition. High tolerances are likely to not satisfy this condition, resulting into the line search algorithm changing direction, converging to the initial control state. As shown later in section \ref{sec:4}, the reader should bear in mind that the state equations march forward in time starting from an initial condition at $t=0$, whereas the adjoint equations are posed backwards in time starting from a terminal condition at $t=T$.\\

\subsubsection{Adjoint-based Gradient}
\label{subsubsec:adj_grad}

\noindent The sensitivity equations can be derived by differentiating the state equations with respect to $\phi$, resulting in \begin{equation}
\frac{\partial \mathcal{N}}{\partial Q} \frac{\mathrm{d}Q}{\mathrm{d}\phi} + \frac{\partial \mathcal{N}}{\partial \phi}=0.
\label{eq:grad_1}
\end{equation} Rearranging (\ref{eq:grad_1}) and substituting it into (\ref{eq:grad_0}) leads to \begin{equation}
\frac{D\mathcal{J}}{D\phi} =  \frac{\partial \mathcal{J}}{\partial \phi} - \frac{\partial \mathcal{J}}{\partial Q} \left(\frac{\partial \mathcal{N}}{\partial Q}\right)^{-1} \left(\frac{\partial \mathcal{N}}{\partial \phi}\right),
\label{eq:grad_2}
\end{equation} where the solution to the adjoint system (\ref{eq:def_adj_eq}) is  \begin{equation}
Q^* = \left[\frac{\partial \mathcal{J}}{\partial Q} \left(\frac{\partial \mathcal{N}}{\partial Q}\right)^{-1}\right]^{\text{T}}.
\end{equation} The adjoint equations can now be expressed as \begin{equation}
\left(\frac{\partial \mathcal{N}}{\partial Q}\right)^{\text{T}} Q^* = \left(\frac{\partial \mathcal{J}}{\partial Q}\right)^{\text{T}}.
\end{equation} Its solution can be used to compute the gradient of the functional as \begin{equation}
\frac{D\mathcal{J}}{D\phi} = \frac{\partial \mathcal{J}}{\partial \phi} - \left( Q^*\right)^{\text{T}} \frac{\partial \mathcal{N}}{\partial \phi}.
\label{eq:adj_grad}
\end{equation} Note that for (\ref{eq:adj_grad}), the cost of computing the gradient is independent of the number of control parameters and is only necessary to solve the adjoint system once to obtain the gradients for every control parameter.\\

\noindent There are two different approaches to derive the adjoint of a system, where both of them have their advantages and disadvantages with respect to each other. These two methods are the continuous or \textit{differentiate then discretise}, and the discrete or \textit{discretise then differentiate}. The continuous approach consists of deriving the adjoint equations analytically following (\ref{eq:def_adj_eq}), where they are then discretised with a numerical scheme to be solved numerically. This procedure provides great flexibility in terms of choosing the numerical discretisation. On the other hand, the discrete approach starts from the already discretised forward system which is then differentiated to obtain the discrete adjoint operator. Note that with this method, the numerics used to solve the forward system of equations are kept during the entire derivation. This includes the numerical scheme used, boundary conditions, and so on. Note also that if any of these numerical methods are changed the whole derivation procedure must be repeated. To make this process simpler, automatic differentiation tools \cite{dadjointcharles} or matrix-free methods \cite{schmid_mfree} can be used. The main characteristic of the discrete approach is that the sensitivities and the gradients obtained are consistent with the discretised problem. This means that, for example, in a very under-resolved simulation, the discrete adjoint will provide the `right' gradients to minimise the cost function even though the physics of the problem are corrupted due to poor grid resolution. On the contrary, the continuous adjoint would fail to give the the `correct' gradients to minimise the given cost function. This same situation would occur if the forward equations suffer from modelling errors. Theoretically, both continuous and discrete approaches would converge to the same result when the discretisation errors vanish ($\Delta \vec{x} \rightarrow 0$) and the equations are physically correct. Vishnampet et al. \cite{bodony} claimed a higher `accuracy' in the sense above of their discrete adjoint code against their continuous one when comparing the gradients from both adjoint solvers with a finite difference approach. In their study, they considered a shear-layer in an open flow, where the boundary conditions were of characteristic-type and they also assumed viscosity to be constant. A common practice when dealing with characteristic boundary conditions in continuous adjoint codes is to neglect the boundary terms that arise from the derivation of the continuous adjoint equations (see section \ref{subsubssec:adjcbc}). Similarly, the assumption of constant viscosity might have a different effect in the forward and adjoint continuous equations. In the authors' opinion it is unsurprising that they claimed an accuracy down to numerical precision for the discrete adjoint, as both forward and discrete adjoint systems are the same. Further discussion about the differences of both approaches can be found, for example, in \cite{gunz,giles_adj}.

\begin{figure}
	\centering
	\begin{subfigure}[b]{0.49\textwidth}
		\centering
		\subcaption{}
	    \begin{tikzpicture}[trim axis left, trim axis right]

\begin{axis}[
xlabel={$\longleftarrow t$},
ylabel={$\lVert Q^* \rVert$},
scale only axis,
legend cell align=left,
xmin=0, xmax=53,
x filter/.code={\pgfmathmultiply{#1}{0.1375}},
xtick={0,15,30,45},
xticklabels={$T_0$,$T_0+15$,$T_0+30$,$T_0+45$},
ymode=log,
every axis y label/.style={at={(current axis.west)},rotate=90,above=0.9cm},
xticklabel style={
rotate=0,
/pgf/number format/precision=1,
/pgf/number format/fixed,
/pgf/number format/fixed zerofill,
},
width=6.6cm,
height=2.75cm,
xmajorgrids,
ymajorgrids,
legend  style={fill=none,
        draw=none,
        anchor=south east,
        legend columns=3,
        column sep=5pt,
        at={(1,1.005)}},
]

\addplot [black]
coordinates {
(0,25.4729152073082)
(1,25.9412548352935)
(2,26.4238026611006)
(3,26.9305049921676)
(4,27.4736577030156)
(5,28.0879607420336)
(6,28.8662943991594)
(7,29.9335649788327)
(8,31.3258049460801)
(9,32.8396143282747)
(10,34.1081216018594)
(11,34.950371435176)
(12,35.4304122511254)
(13,35.5758576999758)
(14,35.3616829540637)
(15,34.8132062274394)
(16,33.9836327191436)
(17,32.932702763542)
(18,31.733989809355)
(19,30.4697494045097)
(20,29.2199122264796)
(21,28.0516970072528)
(22,27.0195755853659)
(23,26.1694282934943)
(24,25.5127378208866)
(25,25.0083290856209)
(26,24.597090922709)
(27,24.2436529813003)
(28,23.9493731822456)
(29,23.7500609665102)
(30,23.7047376638755)
(31,23.8733631674217)
(32,24.2832460903961)
(33,24.8876458526743)
(34,25.5465517252149)
(35,26.0908711075883)
(36,26.4563002552385)
(37,26.6927541675519)
(38,26.806670873702)
(39,26.7230849300689)
(40,26.3863293125792)
(41,25.8126547176515)
(42,25.070625106067)
(43,24.2396960265048)
(44,23.3946583926089)
(45,22.6152519414639)
(46,21.9784680588072)
(47,21.5302140873617)
(48,21.2794890988683)
(49,21.2270531035226)
(50,21.3935308469805)
(51,21.8053208499648)
(52,22.4283463404264)
(53,23.089465432508)
(54,23.5936021952343)
(55,23.9629879402481)
(56,24.167235182174)
(57,24.1168710573551)
(58,24.1626036771825)
(59,24.5909447434757)
(60,25.2948168342076)
(61,26.1595771797628)
(62,27.1078987658265)
(63,28.0273256981307)
(64,28.8526893669731)
(65,29.7241508780516)
(66,30.8138601229478)
(67,31.8018211843771)
(68,32.1676610237379)
(69,31.8162458086418)
(70,30.9249176683375)
(71,29.6900526382945)
(72,28.2800179784103)
(73,26.832725417478)
(74,25.4526322739741)
(75,24.2129685730631)
(76,23.1573918336336)
(77,22.300147375797)
(78,21.6285608569301)
(79,21.111453305899)
(80,20.7175808155093)
(81,20.4394484234027)
(82,20.3047310541738)
(83,20.3436468988332)
(84,20.5087819167531)
(85,20.6629308299185)
(86,20.7000793846769)
(87,20.5970306010513)
(88,20.3439595825048)
(89,19.9324660999769)
(90,19.388693348615)
(91,18.762967572018)
(92,18.0883027984254)
(93,17.3725482430314)
(94,16.6338744067092)
(95,15.9141272317236)
(96,15.2537080113009)
(97,14.6740169914566)
(98,14.1829171512407)
(99,13.7839652725746)
(100,13.4840671908364)
(101,13.3084160986914)
(102,13.3078273404845)
(103,13.5203437986227)
(104,13.872469318052)
(105,14.1644022020308)
(106,14.2906464495419)
(107,14.3249003979135)
(108,14.3022637254916)
(109,14.1830006987089)
(110,13.9855066655413)
(111,13.8228490907206)
(112,13.858391028232)
(113,14.1841226568431)
(114,14.6880487017836)
(115,15.164681115192)
(116,15.4949052370445)
(117,15.6670096398143)
(118,15.7230413537048)
(119,15.7075743197053)
(120,15.6027327793516)
(121,15.3585276458226)
(122,14.9572751503897)
(123,14.4199459226356)
(124,13.7858967932562)
(125,13.0989913698448)
(126,12.40152445711)
(127,11.7312867019014)
(128,11.1196015475708)
(129,10.590861918086)
(130,10.1637663718431)
(131,9.85210565118213)
(132,9.66039837847343)
(133,9.57649591897859)
(134,9.57151934076811)
(135,9.60476102475065)
(136,9.62737304806535)
(137,9.60045455549191)
(138,9.53261314604124)
(139,9.47280845921607)
(140,9.44282818407008)
(141,9.43019833030412)
(142,9.43041709734245)
(143,9.4461188935451)
(144,9.45434724109339)
(145,9.41114228057556)
(146,9.30262890462157)
(147,9.16079335606895)
(148,9.02620217153718)
(149,8.91569256091735)
(150,8.82033109682978)
(151,8.71944341442607)
(152,8.5932868609555)
(153,8.43100140816449)
(154,8.23164107518651)
(155,7.99895415414094)
(156,7.74132610535428)
(157,7.47921347063373)
(158,7.24520942067075)
(159,7.0652518122218)
(160,6.93659176442015)
(161,6.83447721242503)
(162,6.73983265724655)
(163,6.66208882430795)
(164,6.62956445521333)
(165,6.66753436235403)
(166,6.73866105425693)
(167,6.73580407051212)
(168,6.6363343255389)
(169,6.52701731327056)
(170,6.45835440492949)
(171,6.43086863024156)
(172,6.41897046650737)
(173,6.38662655839535)
(174,6.3164224140648)
(175,6.21172275833612)
(176,6.08270921222429)
(177,5.93990075336302)
(178,5.7923143690257)
(179,5.6454135285584)
(180,5.50176735954017)
(181,5.36350205521271)
(182,5.23443273189724)
(183,5.12133522263138)
(184,5.03319043868369)
(185,4.97897919941846)
(186,4.96621925578753)
(187,4.99931300970118)
(188,5.07808610426)
(189,5.19431906406397)
(190,5.32797854816399)
(191,5.45085875020387)
(192,5.53982867684769)
(193,5.59487971839915)
(194,5.63973794141125)
(195,5.69275296656285)
(196,5.74848397933164)
(197,5.79165087417525)
(198,5.81143412292857)
(199,5.80316877294517)
(200,5.76641433149193)
(201,5.70237261749569)
(202,5.61269668946554)
(203,5.50085073279077)
(204,5.37030056713172)
(205,5.22114903631154)
(206,5.05148853034757)
(207,4.86220952510614)
(208,4.65972740286119)
(209,4.45472621392081)
(210,4.25751512759433)
(211,4.07382951310048)
(212,3.90843378107)
(213,3.76794830470755)
(214,3.64894392228588)
(215,3.5479435858131)
(216,3.47054409099312)
(217,3.41639366880779)
(218,3.38317505742886)
(219,3.37062656112272)
(220,3.3830223678537)
(221,3.43252879259088)
(222,3.52967702894009)
(223,3.64995278379111)
(224,3.73637123362625)
(225,3.7682981032254)
(226,3.75855512000298)
(227,3.71903072911476)
(228,3.65983934128689)
(229,3.59164019169165)
(230,3.52265575388063)
(231,3.45640771652526)
(232,3.39433778176239)
(233,3.3385268992259)
(234,3.2906133283624)
(235,3.25129009400227)
(236,3.22017638482266)
(237,3.19507055906009)
(238,3.17439547403829)
(239,3.16076472126477)
(240,3.16165911383126)
(241,3.18712391116301)
(242,3.24603339141148)
(243,3.34145071337552)
(244,3.46504074717302)
(245,3.59461051818999)
(246,3.70051364072286)
(247,3.76085777097111)
(248,3.77634508047229)
(249,3.76847872156757)
(250,3.75595944008236)
(251,3.73716371331872)
(252,3.69962252726096)
(253,3.63580980532927)
(254,3.54600034055477)
(255,3.43502831178534)
(256,3.30935654055546)
(257,3.17449335529596)
(258,3.03421032387791)
(259,2.89036610429355)
(260,2.74284585308061)
(261,2.5911765249364)
(262,2.43617562346009)
(263,2.28089193188654)
(264,2.13075940010066)
(265,1.99176033235122)
(266,1.86701113666558)
(267,1.75517494448407)
(268,1.65285731601397)
(269,1.55882770149199)
(270,1.47644048897816)
(271,1.40954302226313)
(272,1.35575418873079)
(273,1.31024603880795)
(274,1.27196193981607)
(275,1.24298398365413)
(276,1.22918173579511)
(277,1.23879690168077)
(278,1.26410393611597)
(279,1.27069457384623)
(280,1.23701711618154)
(281,1.17769636487801)
(282,1.11573428896196)
(283,1.06423914357715)
(284,1.02744121015364)
(285,1.00406633710374)
(286,0.990169020362175)
(287,0.981297067167634)
(288,0.974319251711122)
(289,0.969150874480968)
(290,0.96803276509746)
(291,0.972934594823632)
(292,0.983695692256538)
(293,0.998251687516942)
(294,1.01420744456849)
(295,1.02952562497264)
(296,1.04282620547721)
(297,1.05453859182813)
(298,1.06852977453134)
(299,1.09032817739939)
(300,1.11869852737163)
(301,1.14057534985643)
(302,1.14117279019383)
(303,1.11611091674261)
(304,1.07119428793396)
(305,1.01522138430356)
(306,0.955336221857879)
(307,0.896176795875699)
(308,0.840357975705636)
(309,0.789111180870154)
(310,0.743058320817469)
(311,0.702229225203423)
(312,0.665580717373829)
(313,0.631274347347322)
(314,0.596414500081682)
(315,0.558755221082297)
(316,0.519263419871175)
(317,0.479831366979109)
(318,0.441992030207967)
(319,0.408285760336588)
(320,0.380982051357746)
(321,0.35868099874952)
(322,0.337324496493377)
(323,0.314695119011952)
(324,0.29157952407598)
(325,0.270150831181016)
(326,0.25245823938641)
(327,0.238210125196664)
(328,0.225336381856313)
(329,0.213845036832366)
(330,0.206137065576924)
(331,0.202518218478863)
(332,0.200475686413596)
(333,0.197046758815622)
(334,0.189507611053246)
(335,0.17771295680897)
(336,0.166222101341762)
(337,0.160114618161596)
(338,0.160458337210359)
(339,0.164643521406028)
(340,0.167831369682046)
(341,0.164826422529716)
(342,0.153345028432216)
(343,0.136786013354116)
(344,0.121849233439143)
(345,0.112491912556268)
(346,0.107046248656414)
(347,0.101903048157721)
(348,0.0961950409023956)
(349,0.090688506615438)
(350,0.0868523353554344)
(351,0.0850168147944338)
(352,0.0835710910251248)
(353,0.082197371529799)
(354,0.0816531750895488)
(355,0.0816438910250447)
(356,0.0832065796346557)
(357,0.0854345835599259)
(358,0.0861855136558566)
(359,0.085473409277603)
(360,0.0840046476191589)
(361,0.082815992026819)
(362,0.0790941616350335)
(363,0.0729468298147538)
(364,0.0704335831975415)
(365,0.0713072484553582)
(366,0.0728120485986906)
(367,0.0744425021719658)
(368,0.0727741971734791)
(369,0.0665930863790008)
(370,0.056485014839502)
(371,0.0469857323990465)
(372,0.0405646477265239)
(373,0.0346822698698895)
(374,0.0273269786987378)
(375,0.019128405837432)
(376,0.0116066968544581)
(377,0.00703354933487853)
(378,0.00368166244391935)
(379,0.00236887285005836)
(380,0)

};

\end{axis}

\end{tikzpicture}
	    \label{fig:adj_div}
	\end{subfigure}
	~
	\centering
	\begin{subfigure}[b]{0.49\textwidth}
		\subcaption{}
		\centering
	    \definecolor{c808080}{RGB}{128,128,128}
\definecolor{cffccaa}{RGB}{255,204,170}
\definecolor{cb3b3b3}{RGB}{179,179,179}

\begin{tikzpicture}[trim axis left, trim axis right]
			\begin{axis}[
			enlargelimits=false,axis on top,
			scale only axis,
			xlabel={$x$},
			ylabel={$y$},
			every axis y label/.style={at={(current axis.west)},rotate=90,above=0.40cm},
			width=6.6cm,
			height=2.75cm,
			xmin=-1.5,xmax=4.5,
			ymin=0,ymax=2.5
			]
			\addplot graphics [xmin=-1.5,xmax=4.5,ymin=0,ymax=2.5]		{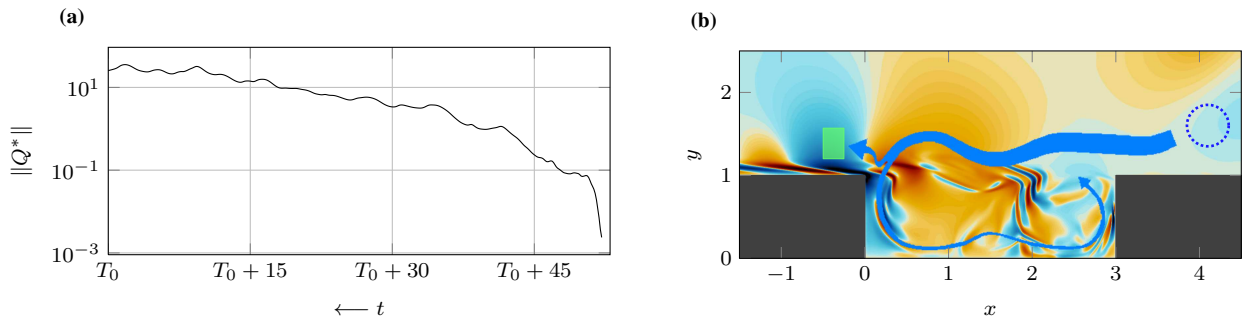};
             
			\addplot[color=blue!90,
			 densely dotted,
            samples=51,
            domain=0:2*pi,
            line width=1pt]
            ({4.1+0.25*cos(deg(x))}, 
             {1.6-0.25*sin(deg(x))});   

             \addplot[patch,patch type=rectangle,green!60,opacity=0.7,shader=flat]
    		 coordinates { (-0.5,1.20) (-0.25,1.20) (-0.25,1.57) (-0.5,1.57)};          
             
			 \addplot[patch,patch type=rectangle,c808080!50!black,shader=flat]
    		 coordinates { (-1.5,0) (0,0) (0,1) (-1.5,1)};
    		 \addplot[patch,patch type=rectangle,c808080!50!black,shader=flat]
    		 coordinates { (3,0) (4.5,0) (4.5,1) (3,1)};    
    
    \end{axis}
\end{tikzpicture}
	    \label{fig:adj_field}
	\end{subfigure}
	\caption{\textbf{(a)} Norm of the adjoint-field for the current 2D cavity. Note that adjoint simulations run backwards in time starting from a terminal condition. \textbf{(b)} Sketch of the gradient computation setup with an instantaneous snapshot of the adjoint-field ($m_v^*$). The green box and the blue dashed circle depict the flow actuation and sensing regions, respectively.}
\label{fig:adjoint_setup}
\vspace{-0.5cm}
\end{figure}

\subsubsection{Limitations of Adjoint-based Optimisation: Model Predictive Control}
\label{subsec:mpc}

\noindent In a separated flow, the trajectories of two neighbouring fluid particles diverge exponentially with time due to the chaotic behaviour of turbulence. Consequently, the initial condition is `forgotten' by the system and the system's sensitivities cannot be computed because the initial and final conditions are unrelated. This means that if the prediction time horizon is longer than the time it takes two neighbouring trajectories to diverge, the computed sensitivities will be incorrect. Lea et al. \cite{adj_chaos_1} used the Lorenz attractor as an example of chaotic systems, showing how the adjoint analysis failed to give the correct time-averaged sensitivities. They suggested that for relatively long time horizons with respect to the predictability of the system's time scales, the adjoint quantities diverge exponentially due to cumulative error growth (figure \ref{fig:adj_div}). Hence, the use of techniques such as receding horizon \cite{mpc_review} are necessary.\\

\noindent Bewley et al. \cite{bewley} carried out flow control on a turbulent channel flow, where a larger time horizon was shown to be beneficial when controlling time-averaged flow quantities. On the other hand, they stated that as the prediction horizon is increased, the required effort to optimise the dynamical system for a given control scheme increases exponentially. An intrinsic limitation in the choice of this time horizon $T$ resides in the travelling time of the control flow scales from the actuation location to the target region, which depends exclusively on the flow characteristics between these two locations (figure \ref{fig:flow_setup_b}). Note that the same limitation occurs in adjoint simulations in the opposite direction (figure \ref{fig:adj_field}).


\section{Flow Simulations}
\label{sec:3}
\subsection{Governing Equations}
\label{subsec:3_1}

The full compressible Navier-Stokes equations govern the fluid flow in our simulations. This set of equations are derived by applying the mass conservation, Newton's second law (momentum conservation) and the conservation of energy, resulting into a non-linear set of five partial differential equations (PDEs), known as continuity (\ref{eq:conteq}), momentum (\ref{eq:momeq}) and energy (\ref{eq:enereq}) equations, which are defined for the entire flow domain $\Omega$. The equations are used in their conservative form, where the variables are density $\rho$, the three-dimensional velocity vector $\vec{u}$ and the total energy $E=T/\left[\gamma\left(\gamma-1\right)M_\infty^2\right]+\frac{1}{2}\vec{u} \cdot \vec{u}^{\text{T}}$,
\begin{gather}
\label{eq:conteq}
\frac{\partial \rho}{\partial t} + \nabla \cdot \left(\rho\vec{u}\right)=0 \\
\label{eq:momeq}
\frac{\partial \rho \vec{u}}{\partial t} + \nabla \cdot \left(\rho \vec{u} \otimes \vec{u} \right) = -\nabla p +\nabla \cdot \tau +\rho \vec{f} \\
\label{eq:enereq}
\frac{\partial \rho E}{\partial t} + \nabla \cdot \left(\rho E \vec{u}\right) = -\nabla \cdot \left(p \vec{u}\right) + \nabla \cdot \left(\tau \cdot \vec{u}\right) + \nabla \cdot \vec{q} +\rho \vec{u} \cdot \vec{f},
\end{gather} 
where the vector $\vec{f}$ represents an external body force. Since the flow is considered compressible, the equation of state (\ref{eq:eos}) gives closure to the system linking pressure $p$ with density $\rho$ and temperature $T$,\begin{equation}
\label{eq:eos}
p = \frac{\rho T}{\gamma M_\infty^2}.
\end{equation} 
The shear stress $\tau$ tensor is symmetric and is defined as \begin{equation}
\tau = \frac{\mu}{Re_\infty}\left(\nabla\vec{u}+\left(\nabla\vec{u}\right)^{\text{T}}\right) - \left(\frac{2}{3}\frac{\mu}{Re_\infty}\nabla \cdot \vec{u}\right) \delta,
\end{equation} with $\mu$ as dynamic viscosity and $\delta$ as the identity matrix. The term $\vec{q}$ in (\ref{eq:enereq}) is known as heat flux vector, which is defined as \begin{equation}
\vec{q} = \frac{\mu}{Re_\infty \left(\gamma-1\right)M_\infty^2Pr_\infty}\nabla T,
\end{equation} where the Prandtl number is assumed to be constant at $Pr=0.72$, and $\gamma=1.4$. The molecular viscosity $\mu$ is computed using Sutherland's
law \cite{white}, setting the ratio of the Sutherland constant over free-stream temperature to $0.36867$.\\

\subsection{Numerical Method}
\label{subsec:3_2}
   
The software used to carry out the numerical simulations is a parallel in-house FORTRAN code called HiPSTAR (High-Performance Solver for Turbulence and Aeroacoustics Research; \cite{rsand}), which solves the governing equations by using generalized curvilinear coordinates. If running three-dimensional simulations, the code assumes periodicity in the spanwise direction, hence this direction is discretised using Fourier transformations (FFTW library \cite{FFTW}). The streamwise and vertical directions are discretised with an explicit fourth-order accurate standard central finite difference scheme. At the boundaries, one-sided explicit schemes are used \cite{carpenter}, being as accurate as the interior scheme. The solution advances in time with an explicit ultra-low storage Runge-Kutta method, which achieves fourth order accuracy with a five step method \cite{RKmethod}. To enhance the stability of the numerical scheme, skew-symmetric splitting of the non-linear terms present in the forward equations is applied using the method suggested by Kennedy and Gruber \cite{skewsplit}. A sixth-order accurate explicit filter \cite{exp_filter} with a weighting of 0.2 is applied after every Runge-Kutta cycle to remove the spurious numerical high-wavenumber oscillations. The parallelisation is achieved with a combined method of MPI calls in the streamwise and vertical directions with OMP threads in the spanwise direction.\\

\noindent Open boundaries are treated with characteristic boundary conditions (CBC; \cite{thompson1,thompson2}), which are used to avoid spurious numerical reflections from the boundaries. An integral formulation of CBC \cite{jones} is applied at the inflow boundary, whereas a zonal CBC \cite{zonalcbc} is used at the outflow. Wall boundaries are modelled as isothermal no-slip walls where the conditions imposed are
\begin{equation}
u_w=v_w=w_w=0, \hspace{0.5cm} T_w=const.
\end{equation}


\section{Adjoint Simulations}
\label{sec:4}

As introduced in section \ref{subsec:adj_method}, there are two main approaches to tackle the adjoint problem, the \textit{discrete} and the \textit{continuous} approaches. One of the fundamental sources of disagreement between both methods resides in the resulting gradient of the cost functional as the spatial resolution decreases. Direct Numerical Simulations (DNS) are carried out in the current research, which implies that no modelling assumptions are considered, and also the temporal and spatial discretisations are sufficiently fine to capture all the events contained in the flow field. Given these conditions, the discrepancies between the continuous and discrete approaches should be minimal \cite{cont_disc_adj}. For convenience and flexibility in the implementation, the continuous approach is chosen in the present investigation.

\subsection{Adjoint Equations}
\label{subsec:4_1}

The present adjoint formulation uses the energy equation (\ref{eq:enereq}) rearranged in terms of pressure. This is done by deriving the internal energy equation and then substituting in the equation of state. The internal energy equation is obtained by subtracting the kinetic energy equation from the total energy equation. The kinetic energy is calculated as \begin{equation}
\label{eq:rhoekin}
\frac{\partial \rho e_{kin}}{\partial t} = \vec{u} \cdot \mathrm{momentum} -\frac{1}{2} \vec{u}^2 \cdot \mathrm{continuity},
\end{equation}
where the derivatives can be summed up using the product rule. Hence, subtracting equation (\ref{eq:rhoekin}) from the total energy equation and using the equation of state (\ref{eq:eos}) gives
\begin{equation}
\label{eq:pforw}
\frac{\partial p}{\partial t} + \nabla \cdot \left(p\vec{u}\right) + \left(\gamma -1 \right) \left[ p \left(\nabla \cdot \vec{u}\right) + \nabla \cdot \vec{q} + tr\left(\tau \cdot \nabla \vec{u} \right)   \right]=0.
\end{equation}

\noindent Following the definition given in (\ref{eq:def_adj_eq}), the adjoint equations arise from the linearisation of eqs. (\ref{eq:conteq}), (\ref{eq:momeq}) and (\ref{eq:pforw}) as
\begin{equation}
\lim_{\epsilon \to 0} \left(\frac{\mathcal{L}\left(Q+\epsilon \tilde{Q},Q^*,\phi \right) - \mathcal{L}\left(Q,Q^*,\phi \right)}{\epsilon}\right) = 0,
\label{eq:LFuncDiff}
\end{equation}
where $\tilde{Q}$ is an arbitrary state. As the equations considered here are partial differential equations, integration by parts is required to remove, if possible, the derivatives from the arbitrary state $\tilde{Q}$. This last step introduces additional terms which are only defined at the boundary of the domain $\Gamma$, known as \textit{boundary terms}. Note that (\ref{eq:LFuncDiff}) satisfies the Green-Lagrange identity \begin{equation}
\left< Q^*,\mathcal{N}'\left(Q,\phi\right)\tilde{Q} \right> = \left< \mathcal{N}^*\left(Q,\phi\right)Q^*,\tilde{Q} \right> + BT,
\label{eq:glid}
\end{equation} where $BT$ represents the boundary terms. These boundary terms should not be neglected since they determine the correct boundary conditions of the adjoint problem.

\subsubsection{Adjoint Euler Equations}

The state variables chosen for the derivation of the adjoint governing equations are $Q = \left[ p,\rho u,\rho v, \rho w, \rho \right]$, where their corresponding co-state or adjoint variables are $Q^*=\left[\rho^*,m_u^*,m_v^*,m_w^*,p^*\right]$. For the sake of clarity, the viscous terms of the governing equations are grouped as the RHS $Sv$ terms for each equation. Hence, the adjoint Euler equations can be written as
\begin{gather}
\label{eq:adjointEuler1}
-\frac{\partial \rho *}{\partial t} - \vec{u} \cdot \nabla \rho^* + \left(\gamma - 1\right)\rho^* \nabla \cdot \vec{u} - \nabla \cdot \vec{m}^* = Sv_{\rho^*} \\
\label{eq:adjointEuler2}
-\frac{\partial \vec{m}^*}{\partial t} - \gamma\frac{p}{\rho} \nabla \rho^* - \left(\gamma-1\right)\frac{\rho^*}{\rho}\nabla p - \vec{u}\left(\nabla \vec{m}^*+\left(\nabla \vec{m}^*\right)^T\right) - \nabla p^* = Sv_{\vec{m}^*}\\
\label{eq:adjointEuler3}
-\frac{\partial p^*}{\partial t} + \gamma\frac{p\vec{u}}{\rho} \cdot \nabla\rho^* + \left(\gamma -1 \right)\frac{\rho^*}{\rho}\vec{u}\cdot \nabla p + \vec{u} \cdot \left(\vec{u}\cdot \nabla \right)\vec{m}^* = Sv_{p^*}.
\end{gather}

\noindent Note that the derivation of the above adjoint equations relates the adjoint density with pressure and the adjoint pressure with density \cite{bewley2}. Hence, both forward and adjoint equations present a ``logical zero-Mach-number limit", where in case of constant $\rho$ and $\rho^*$ the $p$ and $p^*$ equations cancel out. A detailed derivation of these equations can be found in \cite{bewley}.

\subsubsection{Adjoint Navier-Stokes Equations}
\label{subsub:adjns}
In previous studies based on Lagrangian optimisation, the derivation of the adjoint viscous terms has usually been carried out under the assumptions of constant viscosity and negligible viscous dissipation terms in the pressure equation \cite{spagnoli}. On the other hand there is no guarantee that these assumptions have the same effect in the forward and adjoint equations, which might contribute to an inconsistent adjoint field. Hence, no assumptions were made during the derivation of the viscous terms in the present study. The viscous terms of eqs. (\ref{eq:adjointEuler1}), (\ref{eq:adjointEuler2}) and (\ref{eq:adjointEuler3}) are
\begin{gather}
\label{eq:adjointNS1}
Sv_{\rho^*} = \vec{q}^* + \mu^*\chi  \\
\label{eq:adjointNS2}
Sv_{\vec{m}^*} = \frac{1}{\rho} \nabla \cdot \sigma^* \\
\label{eq:adjointNS3}
Sv_{p^*} = -\frac{\vec{u}}{\rho}\cdot \nabla\cdot \sigma^* -\frac{p}{\rho}\left( \vec{q}^* +\mu^*\chi \right),
\end{gather}
where $\vec{q}^*$ and $\sigma^*$ are
\begin{equation}
\vec{q}^* = \frac{\gamma}{Pr_\infty Re_\infty \rho} \nabla \cdot \left(\mu \nabla \rho^*\right),\quad \sigma^* =  \tau^* - 2\left(\gamma - 1\right)\rho^* \tau.
\end{equation}
As the adjoint simulation propagates information in a backward sense, energy must be transmitted from small to bigger structures, hence $\sigma^*$ must consist of a combination of dissipation and production terms. The dissipation is accounted for with 
\begin{equation}
\tau^* = \frac{\mu}{Re_\infty}\left(\nabla\vec{m}^*+\left(\nabla\vec{m}^*\right)^{\text{T}}\right) - \left(\frac{2}{3}\frac{\mu}{Re_\infty}\nabla\cdot\vec{m}^*\right)\delta,
\end{equation}
which could be seen as an adjoint momentum shear stress tensor, whereas the production term depends on the forward shear stress tensor $\tau$. Note that if the viscous dissipation term is neglected in (\ref{eq:pforw}) this production term would vanish. For simplicity, the terms that arise from considering varying viscosity have been collected into the adjoint viscosity $\left(\mu^*\right)$, which leads to the following algebraic equation closing the full adjoint Navier-Stokes equations
\begin{equation}
\label{eq:adjmu}
\mu^* = tr\left(\left(\nabla \vec{m}^*-\left(\gamma -1\right)\rho^*\nabla \vec{u}\right) \cdot \left( \frac{\tau}{\mu}\right)\right) + \frac{1}{PrM_\infty^2Re_\infty}\left(\nabla\rho^*\cdot\left(\nabla T\right)^{\text{T}}\right).
\end{equation}
The adjoint viscosity term is weighted in (\ref{eq:adjointNS1}) and (\ref{eq:adjointNS3}) by $\chi$, which follows directly from assuming the variations in viscosity according to the Sutherland's law, giving
\begin{equation}
\chi = \frac{T^{3/2}}{p} \left(\frac{1+R_{Su}}{T+R_{Su}}\right) \left( \frac{T}{\left(T+R_{Su}\right)} - \frac{3}{2} \right).
\end{equation}

\subsection{Adjoint Boundary Conditions}
\label{subsec:4_2}

\noindent The derivation of appropriate boundary conditions (temporal and spatial) for a continuous adjoint approach starts by accounting for the forward boundary conditions in the Lagrange functional. This will introduce additional adjoint variables $\left(\psi\right)$ which are only defined at the boundaries and they are uncoupled from the adjoint governing equations
\begin{equation}
\mathcal{L} \left(Q,Q^*,\phi \right) = \mathcal{J} \left( Q, \phi \right) - {Q^*}^{\text{T}} \cdot N\left(Q,\phi \right) - \psi^{\text{T}} \cdot BC\left(Q,\phi\right).
\label{eq:LFuncbc}
\end{equation}

\noindent Hence, additional boundary terms are also introduced, closing the algebraic system of equations formed by the boundary terms, whose solution is the adjoint boundary conditions.\\

\noindent As the adjoint equations are posed backwards in time, the adjoint initial condition or `terminal condition' is required to start the simulation. If no specific term is included in the Lagrangian functional to impose a given terminal condition, the adjoint variables are initialized as $\rho^*|_{t=T}=0$, $\vec{m}^*|_{t=T}=0$ and $p^*|_{t=T}=0$, the `null condition'.\\

\noindent The adjoint isothermal no-slip wall conditions are obtained after substituting forward wall boundary conditions into (\ref{eq:LFuncbc}) and solving the system of equations of the full viscous boundary terms. This leads to 
\begin{equation}
\left(m_u^*\right)_w=\left(m_v^*\right)_w=\left(m_w^*\right)_w=\left(\rho^*\right)_w=0.
\end{equation}

\subsubsection{Adjoint Characteristic Boundary Conditions}
\label{subsubssec:adjcbc}

\noindent To achieve an equivalent treatment as in the forward simulations, the free-stream, inflow and outflow boundaries, are also modelled as characteristic  boundary conditions. The idea behind the method of characteristics is to transform the derivatives that are normal to the computational boundary into combinations of the amplitudes of the characteristic waves $\mathscr{L}$, where each one of them has its own characteristic velocity $\lambda$. The derivation of the adjoint characteristic equations in the $\xi$ direction for a uniform Cartesian grid can be found in the appendix \ref{sec:App_a}. With the equations in this form it is relatively straightforward to model a `non-reflecting boundary'. This is done by setting to zero the amplitude of the waves $\mathscr{L}$ whose characteristic velocity $\lambda$ propagates into the domain. Similarly to the forward simulations, this boundary condition results in an ill-posed problem where the quantities would drift with time marching \cite{poinsotlele}. Hence, this condition is only suitable for free-stream boundaries.\\

\noindent If the terminal condition of the adjoint simulation is the null condition, one could model the far-stream or infinity conditions $\left(Q^*_\infty\right)$ as if all the quantities far away from the computational domain are set to zero. Hence, equivalent local one-dimensional inviscid (LODI) relations \cite{poinsotlele} can be derived at the inflow and outflow boundaries, leading to suitable boundary conditions that will prevent a drift of the adjoint quantities. The boundary conditions for inflow and outflow as streamwise boundaries follow from (\ref{chareq_xi}), where after assuming the flow at the boundaries to be locally one-dimensional and inviscid the source term $C_{\xi}$ can be neglected giving the LODI system
\begin{equation}
-\frac{\partial Q^*}{\partial t} + S_{\xi}\mathscr{L} = 0.
\end{equation}

\noindent Solving the above system of equations for $\mathscr{L}$ the adjoint LODI relations can be written as 
\begin{align}
\label{lodi1}
\mathscr{L}_1 =& \frac{\partial p^*}{\partial t} + u \frac{\partial m^*_u}{\partial t} \\
\label{lodi2}
\mathscr{L}_2 =& \frac{\partial m^*_w}{\partial t} \\
\label{lodi3}
\mathscr{L}_3 =& \frac{\partial m^*_v}{\partial t} \\
\label{lodi4}
\begin{split}
\mathscr{L}_4 =& \frac{1}{2c} \left(- \frac{\partial m^*_u}{\partial t} -\frac{\partial p^*}{\partial t}-\left(c^2 \frac{\partial \rho^*}{\partial t}+\frac{\partial m^*_u}{\partial t} u+\frac{\partial m^*_v}{\partial t} v+\frac{\partial m^*_w}{\partial t} w\right)  \right)
\end{split} \\
\label{lodi5}
\begin{split}
\mathscr{L}_5 =& \frac{1}{2c} \left(-\frac{\partial m^*_u}{\partial t} +\frac{\partial p^*}{\partial t} +\left(c^2 \frac{\partial \rho^*}{\partial t}+\frac{\partial m^*_u}{\partial t} u+\frac{\partial m^*_v}{\partial t} v+\frac{\partial m^*_w}{\partial t} w\right)\right),
\end{split}
\end{align}
where each time derivative can be replaced by 
\begin{equation}
\frac{\partial Q^*}{\partial t} = c\left(1-M^2\right)  \sigma_{in/out} \left[Q^* - Q^*_\infty\right].
\end{equation}
The value of $\sigma_{in/out}$ determines the strength of the incoming waves and is chosen to be 0.25 as recommended by \cite{poinsotlele} for the Navier-Stokes equations. These boundary conditions ensure a well-posed system, where for the case of an outflow (adjoint inflow) the values of $\mathscr{L}_1$, $\mathscr{L}_2$, $\mathscr{L}_3$ and $\mathscr{L}_4$ are determined by (\ref{lodi1}) to (\ref{lodi4}), and in case of an inflow (adjoint outflow) $\mathscr{L}_5$ is only modified by (\ref{lodi5}).


\section{Open Cavity Noise Control}
\label{sec:5}

The cavity flow considered in this article has a laminar inflow with Reynolds number $Re = 5000$, based on the cavity height and free-stream velocity. The cavity height is also chosen to non-dimensionalise the stramwise and vertical directions. With the coordinate origin at the lower-left corner of the cavity, the domain ranges from -20 to 20 in $\xi$ and from 0 to 10 in $\eta$. The grid consists of a total of 480000 points, where the DNS resolution was only kept in the vicinity of the cavity. The boundary layer thickness is $\delta \approx 0.28$ at the separation point, where the non-dimensional free-stream velocity is $U_\infty = 1$. The Prandtl and Mach numbers are respectively $Pr_\infty = 0.72$ and $M_\infty = 0.5$. To carry out the adjoint simulations, the flow field is stored every 250 time-steps (at intervals of 0.1375 time units) and it is linearly interpolated in between for every sub-step of the Runge-Kutta scheme. Note that this implicitly decreases the Nyquist frequency of our sensor, acting as a low-pass filter.\\

\begin{figure}[t]
	\centering
    \definecolor{c808080}{RGB}{128,128,128}
\definecolor{cffccaa}{RGB}{255,204,170}
\definecolor{cb3b3b3}{RGB}{179,179,179}

\begin{tikzpicture}[trim axis left, trim axis right]
			\begin{axis}[
			enlargelimits=false,axis on top,
			scale only axis,
			xlabel={$x$},
			ylabel={$y$},
			every axis y label/.style={at={(current axis.west)},rotate=90,above=0.75cm},
			width=11cm,
			height=4cm,
			xmin=-4,xmax=7,
			ymin=0,ymax=4
			]
			\addplot graphics [xmin=-4,xmax=7,ymin=0,ymax=4]		{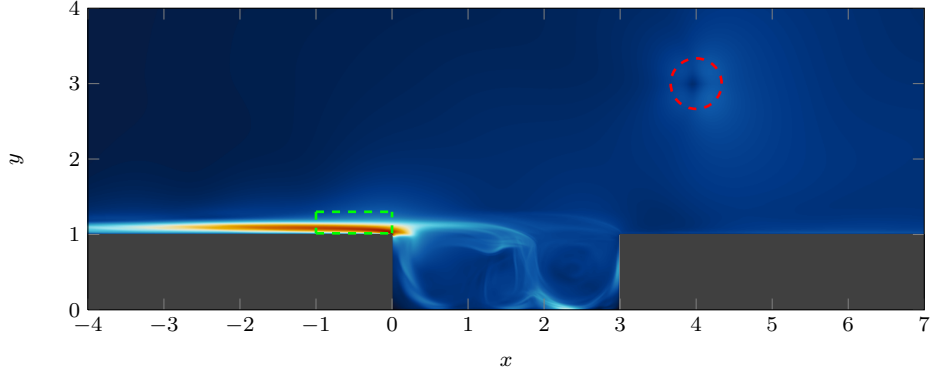};     
			
			 \addplot[patch,patch type=rectangle,c808080!50!black,shader=flat]
    		 coordinates { (-4,0) (0,0) (0,1) (-4,1)};
    		 \addplot[patch,patch type=rectangle,c808080!50!black,shader=flat]
    		 coordinates { (3,0) (7,0) (7,1) (3,1)};    
    		 
			 \addplot [green,dashed, line width = 1pt]
			 coordinates   {(-1,1.015) (0,1.015)};
			 \addplot [green,dashed, line width = 1pt]
			 coordinates   {(-1,1.3) (0,1.30)};
 			 \addplot [green,dashed, line width = 1pt]
			 coordinates   {(-1,1.3) (-1,1.015)};
 			 \addplot [green,dashed, line width = 1pt]
			 coordinates   {(0,1.015) (0,1.30)};
    
            \draw [dashed,red,line width=1pt] (axis cs:4.0,3) circle[radius=0.335cm];
    
    \end{axis}
\end{tikzpicture}
\caption{Root mean square of the magnitude of the adjoint momentum $\vec{m}^*$. Red and blue contours indicate high and low values, respectively. The green-dashed rectangle indicates the actuating sub-volume. The red-dashed circle represents the sensing region.}
\label{fig:adj_rms}
\vspace{-0.5cm}
\end{figure}

\noindent To cost the noise radiation from the trailing edge of the cavity, a convenient cost function for this particular problem is \begin{equation}
\mathcal{J}\left(Q,\phi\right) = \frac{1}{T} \int_0^T\int_{\Omega}\frac{1}{2}\left(p - \overline{p}\right)^2W_s\left(\vec{x}\right)\mathrm{d}\Omega\mathrm{d}t,
\end{equation} where $T$ represents the optimisation horizon and $\overline{p}$ is the averaged pressure over this time period.  $W_s\left(\vec{x}\right)$ is a spatial weighting function that represents the sensing region, which is defined as \begin{equation}
W_s\left(\vec{x}\right) = \mathrm{e}^{-\frac{\left(x-4\right)^2+\left(y-3\right)^2}{0.05}}.
\end{equation} To exploit the advantages of the adjoint method, the actuation is chosen to be a time-dependent body forcing sub-domain $\vec{f}\left(\vec{x},t\right)$, where every grid point within the sub-domain has two control parameters (the streamwise and vertical components of the forcing vector $\vec{f}$). Including both cost function and forcing terms in the Lagrangian functional (\ref{eq:LFunc}), which are then substituted into (\ref{eq:LFuncDiff}), leads to their respective adjoint terms which can be written as \begin{align}
-\frac{\partial \rho^*}{\partial t} &= RHS + \underbrace{\left(p - \overline{p}\right) W_s\left(\vec{x}\right)}_{\text{Cost Function}} \\
-\frac{\partial p^*}{\partial t} &= RHS + \underbrace{\left(\vec{m}^* \cdot \vec{f}\right) W_a\left(\vec{x}\right)}_{\text{Forcing Terms}}.
\end{align} From (\ref{eq:adj_grad}) the gradient for each of the control parameters follows as \begin{equation}
\frac{D\mathcal{J}}{D\phi_{\xi}} = \rho m_u^* \quad \text{and} \quad \frac{D\mathcal{J}}{D\phi_{\eta}} = \rho m_v^*.
\end{equation} To choose an appropriate location for the actuation, an initial statistical study of the adjoint quantities is carried out for an entire time horizon $T$, which gives a measure of the flow sensitivities. Given the time-dependency of the chosen actuation, its control authority should be proportional to the temporal fluctuations of $\vec{m}^*$. Figure \ref{fig:adj_rms} reveals that the area with the highest root mean square values in $\vec{m}^*$ is in the boundary layer before the flow separation point, reaching maximum values right after the leading edge of the cavity. Hence, the location of our actuation is defined as \begin{equation}
W_a\left(\vec{x}\right) = \begin{cases}
	1 & \quad \text{if} \ \vec{x} \in \left[\left(-1,0\right),\left(1.015,1.3\right)\right]\\
	0 & \quad \text{if} \ \vec{x} \notin \left[\left(-1,0\right),\left(1.015,1.3\right)\right]\\
\end{cases}.
\end{equation} The actuating sub-domain contains a total of 8000 grid points. In a similar fashion as for the flow variables in the adjoint simulations, the control parameters are updated every 250 time iterations and they are linearly interpolated in between every capture. Note that this also removes implicitly the high frequencies from the controller. The fact that the sensing region is located downstream of the cavity, in a region of the flow with strong convection, reduces the communication time between the sensor and the actuation. Based on the growth of the adjoint field shown in figure \ref{fig:adj_div} and a preliminary spectral analysis of the pressure fluctuations at the sensor location, the receding horizon is chosen to be $T=19.25$ time units, and the leap forward after the optimisation over the time horizon has converged is $T_a = 8.25$ time units. The optimisation is run for 5 complete horizons, which covers a total of 52.25 time units. This results in a total of 2.256 million control parameters per receding horizon T, and 6.096 million control parameters over the five total optimisation periods.

\subsection{Optimisation Results}

\begin{figure}
	\centering
	\begin{subfigure}[b]{0.49\textwidth}
		\centering
		\subcaption{}
%
%
%
%
\begin{tikzpicture}[trim axis left, trim axis right]

\begin{axis}[
xlabel={Iterations},
ylabel={$\mathcal{J}\left(Q,\phi\right)$},
scale only axis,
legend cell align=left,
xmin=1, xmax=30,
ymax = 2e-5,
x filter/.code={\pgfmathmultiply{#1+1}{1}},
every axis y label/.style={at={(current axis.west)},rotate=90,above=0.9cm},
ymode=log,
width=6cm,
height=4.5cm,
xmajorgrids,
ymajorgrids,
legend  style={fill=none,
        draw=none,
        anchor=north east,
        legend columns=1,
        column sep=5pt,
        at={(1.02,0.98)}},
legend  entries={{$T_1 \hspace{0.5cm} $},{$T_2 \hspace{0.5cm}$},{$T_3 \hspace{0.5cm}$},{$T_4 \hspace{0.5cm}$},{$T_5 \hspace{0.5cm}$}}
]

\addplot [black,solid]
coordinates {
(0,2.3331081e-06)
(1,1.8322796e-06)
(2,1.7000797e-06)
(3,1.5580449e-06)
(4,1.5580449e-06)
(5,1.5282187e-06)
(6,0.0045234468)
(7,1.520771e-06)
(8,1.4869474e-06)
(9,1.4747741e-06)
(10,0.0045248801)
(11,1.4533045e-06)
(12,1.4267439e-06)
(13,1.3881955e-06)
(14,1.2959367e-06)
(15,1.2498352e-06)
(16,1.2335236e-06)
(17,1.1992063e-06)
(18,1.3368132e-06)
(19,1.1909697e-06)
(20,1.1596229e-06)
(21,1.0779509e-06)
(22,1.0519781e-06)
(23,1.0340693e-06)
(24,1.0065337e-06)
(25,1.0065337e-06)

};
\addplot [green!50!black,dashed,line width = 1pt]
coordinates {
(0,6.13823719999999e-07)
(1,4.0045342e-07)
(2,3.56901409999999e-07)
(3,3.07938139999999e-07)
(4,2.94741259999999e-07)
(5,2.88058059999999e-07)
(6,2.7567853e-07)
(7,2.6167385e-07)
(8,2.53259259999999e-07)
(9,2.43709389999999e-07)
(10,2.36588449999999e-07)
(11,2.286429e-07)
(12,2.2522094e-07)
(13,2.2336246e-07)
(14,2.17811299999999e-07)
(15,2.18563319999999e-07)
(16,2.1654219e-07)
(17,2.3207352e-07)
(18,2.17454839999999e-07)
(19,2.1572754e-07)
(20,2.2652655e-07)
(21,2.1563135e-07)
(22,2.1660023e-07)
(23,2.1510053e-07)
(24,2.306323e-07)
(25,2.16483779999999e-07)
(26,2.1378338e-07)
(27,2.1099666e-07)
(28,2.0539184e-07)
(29,2.02335089999999e-07)

};
\addplot [blue,densely dashed ,line width = 1pt]
coordinates {
(0,5.58255859999999e-07)
(1,3.27925509999999e-07)
(2,3.27925509999999e-07)

};
\addplot [red,densely dotted ,line width = 1pt]
coordinates {
(0,5.59112099999999e-07)
(1,4.95982139999999e-07)
(2,3.71007809999999e-07)
(3,3.71007809999999e-07)

};
\addplot [red!60!green,dashdotted ,line width = 1pt]
coordinates {
(0,2.4990845e-07)
(1,3.76737199999999e-07)
(2,2.04561539999999e-07)
(3,1.6652899e-07)
(4,1.6644698e-07)
(5,1.53233e-07)
(6,1.3895041e-07)
(7,1.3561791e-07)
(8,1.3028429e-07)
(9,1.2626222e-07)
(10,1.2150393e-07)
(11,1.2094046e-07)
(12,1.2094046e-07)

};

\end{axis}

\end{tikzpicture}
	    \label{fig:cost_reduction}
	\end{subfigure}
	~
	\centering
	\begin{subfigure}[b]{0.49\textwidth}
		\subcaption{}
		\centering
	    \input{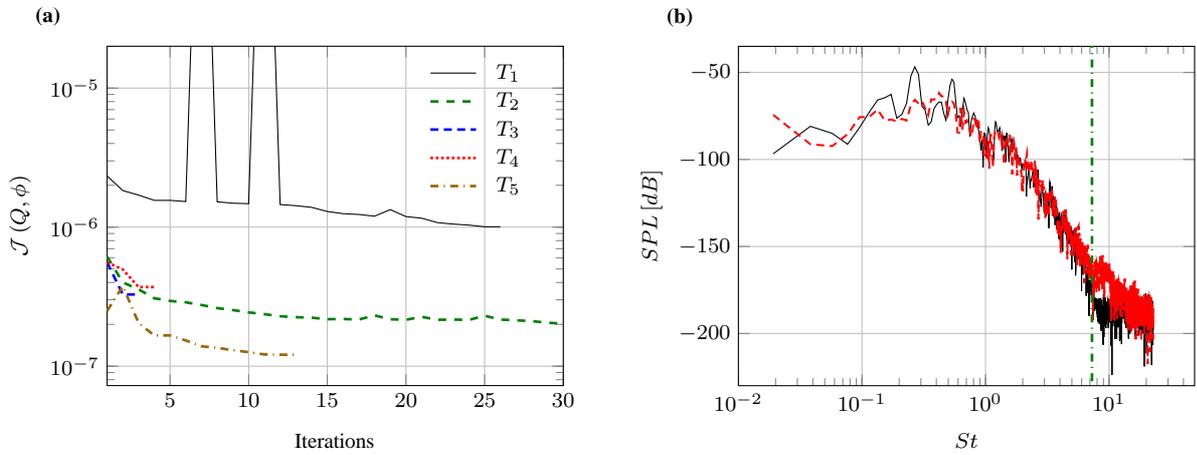}
	    \label{fig:fft_mic}
	\end{subfigure}
	\caption{\textbf{(a)} Value of the cost function for every iteration of the optimisation framework, for all five receding horizons. Note that the required number of iterations to achieve convergence varies for each horizon. \textbf{(b)} Sound pressure level (SPL) at the center of the sensing region for the non-actuated (\ref{line_no_control}) and actuated (\ref{line_control}) simulations. The vertical line  (\ref{c_off_ny}) indicates the flow-capturing frequency, which is also the sampling frequency of the sensor.}
\label{fig:cost_pics}
\vspace{-0.5cm}
\end{figure}

The entire optimisation required a total of 76 function and gradient evaluations, with 4 additional forward simulations to advance the optimisation between receding horizons once the previous one is converged. For the present case, the adjoint computations required a similar amount of time as the forward simulations. Figure \ref{fig:cost_reduction} illustrates the significant reduction in cost for all receding horizons. The local sudden increases in the cost function value (especially in $T_1$) are caused by a poor estimation of the control parameters by the control update algorithm (L-BFGS), which are rapidly corrected in the next iteration. The convergence criterion consisted in the last two function evaluations of every horizon to be equal down to three significant figures, and being the lowest value of the series. Due to limitations in computational resources, the maximum number of functions evaluations per receding horizon was limited to 30.\\

\noindent Once the iterative optimisation has concluded, the effect of the control is evaluated over the entire optimisation horizon by performing controlled and uncontrolled simulations for 52.25 time units each. The reduction achieved in the value of the cost function is $79.6\%$ (without control $\mathcal{J}\left(Q,\phi\right) = 2.679\cdot 10^{-6}$, with control $\mathcal{J}\left(Q,\phi\right) = 5.458\cdot 10^{-7}$). Figure \ref{fig:fft_mic} shows the spectrum of the pressure fluctuations at the center of the sensing region with and without the actuation. The reduction of energy in the lower resonant frequencies is evident, where the amplitude of the resonant frequency located at $St=0.2678$ lessens from $-46.76$dB to $-65.70$dB. Note that there is a considerable increase in the energy present in the high frequencies above the flow-field sampling frequency ($St\approx7.272$). As mentioned earlier, these high frequency fluctuations cannot be captured by the cost function due to an induced aliasing effect, which acts as a low-pass filter in the cost function.\\

\begin{figure}
	\centering
	\begin{subfigure}[b]{0.49\textwidth}
		\centering
		\subcaption{}
	    \definecolor{c808080}{RGB}{128,128,128}
\definecolor{cffccaa}{RGB}{255,204,170}
\definecolor{cb3b3b3}{RGB}{179,179,179}

\begin{tikzpicture}[trim axis left, trim axis right]
			\begin{axis}[
			enlargelimits=false,axis on top,
			scale only axis,
			xlabel={$x$},
			ylabel={$y$},
			every axis y label/.style={at={(current axis.west)},rotate=90,above=0.6cm},
			width=6.6cm,
			height=4.4cm,
			xmin=-1.5,xmax=4.5,
			ymin=0,ymax=4
			]
			\addplot graphics [xmin=-1.5,xmax=4.5,ymin=0,ymax=4]		{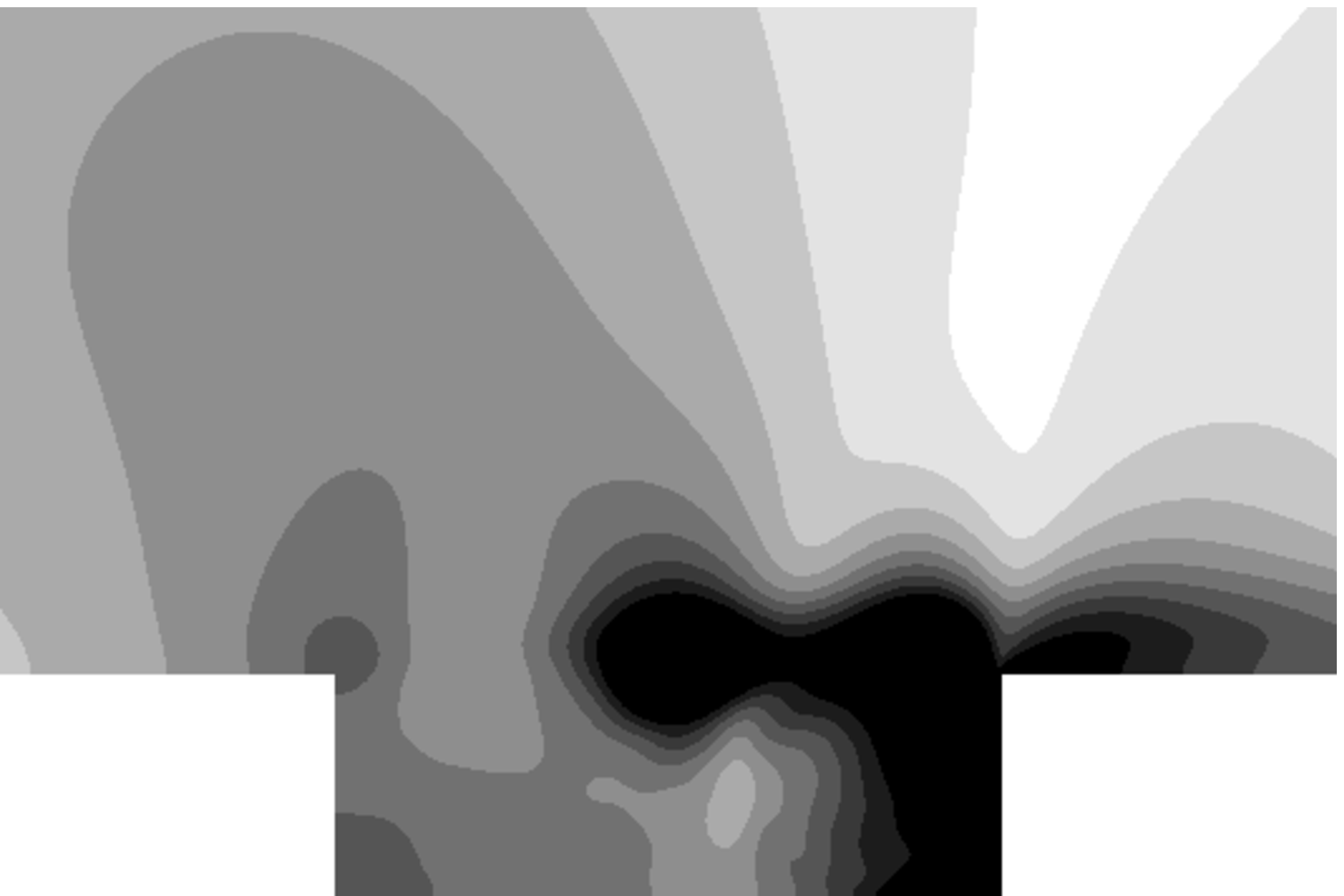};

			 \addplot[patch,patch type=rectangle,c808080!50!black,shader=flat]
    		 coordinates { (-1.5,0) (0,0) (0,1) (-1.5,1)};
    		 \addplot[patch,patch type=rectangle,c808080!50!black,shader=flat]
    		 coordinates { (3,0) (4.5,0) (4.5,1) (3,1)};    
    		 
    		 \draw [dashed,blue,line width=1pt] (axis cs:4.0,3) circle[radius=0.335cm];
    
    \end{axis}
\end{tikzpicture}
	    \label{fig:ncprms}
	\end{subfigure}
	~
	\centering
	\begin{subfigure}[b]{0.49\textwidth}
		\subcaption{}
		\centering
	    \definecolor{c808080}{RGB}{128,128,128}
\definecolor{cffccaa}{RGB}{255,204,170}
\definecolor{cb3b3b3}{RGB}{179,179,179}

\begin{tikzpicture}[trim axis left, trim axis right]
			\begin{axis}[
			enlargelimits=false,axis on top,
			scale only axis,
			xlabel={$x$},
			every axis y label/.style={at={(current axis.west)},rotate=90,above=0.75cm},
			width=6.6cm,
			height=4.4cm,
			xmin=-1.5,xmax=4.5,
			ymin=0,ymax=4
			]
			\addplot graphics [xmin=-1.5,xmax=4.5,ymin=0,ymax=4]		{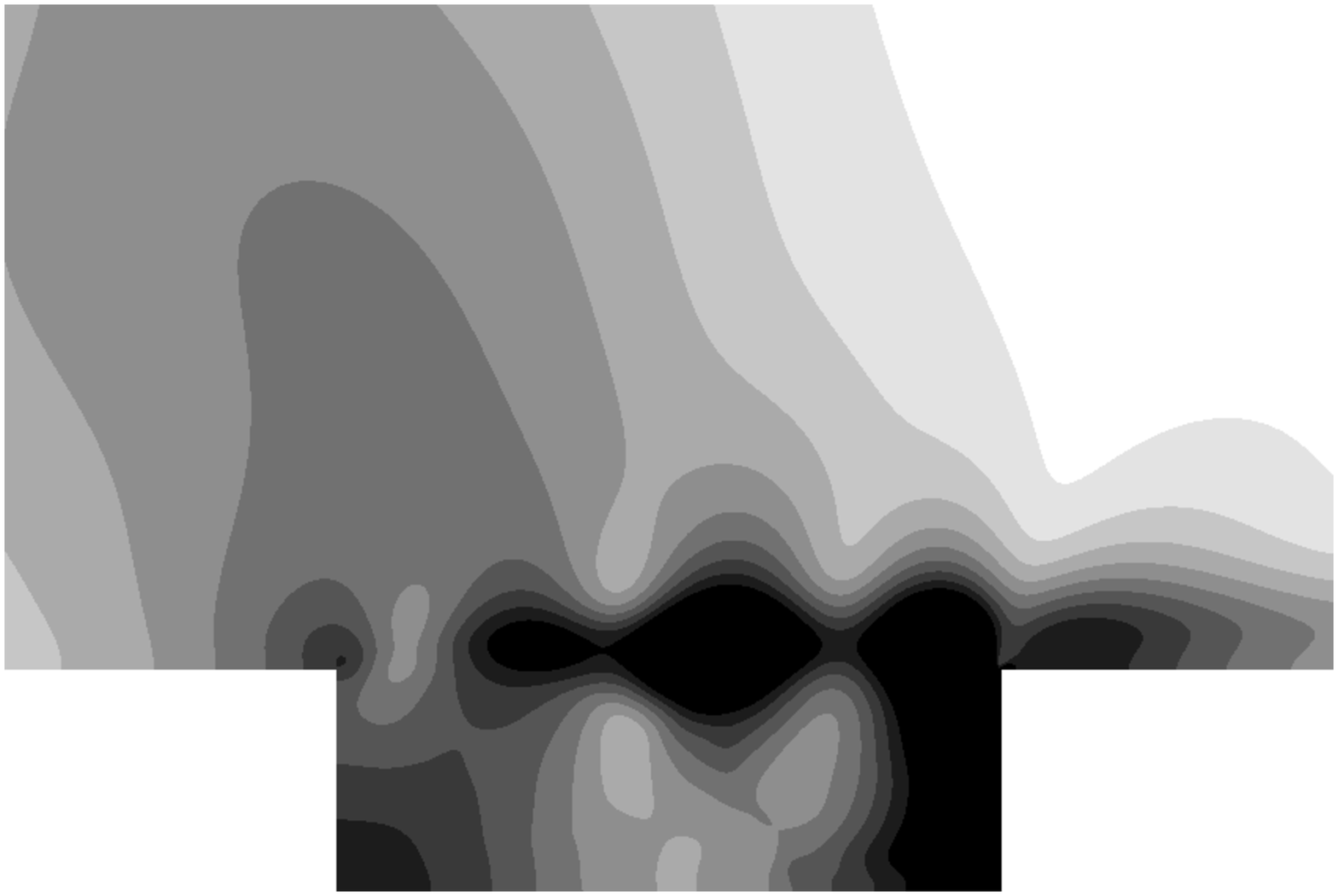};
			
			 \addplot[patch,patch type=rectangle,c808080!50!black,shader=flat]
    		 coordinates { (-1.5,0) (0,0) (0,1) (-1.5,1)};
    		 \addplot[patch,patch type=rectangle,c808080!50!black,shader=flat]
    		 coordinates { (3,0) (4.5,0) (4.5,1) (3,1)};    

    		 \draw [dashed,blue,line width=1pt] (axis cs:4.0,3) circle[radius=0.335cm];    
    
    \end{axis}
\end{tikzpicture}
	    \label{fig:cprms}
	\end{subfigure}
	\caption{Root mean square pressure ($p_{rms}$) contours for uncontrolled \textbf{(a)} and controlled \textbf{(b)} cases, respectively. Sensing region is indicated with a dashed-blue circle. Darker contours represent higher values of $p_{rms}$, ranging from 0 to 0.05.}
\label{fig:prms}
\vspace{-0.5cm}
\end{figure}

\noindent Figures \ref{fig:prms} shows the root mean square of pressure ($p_{rms}$) for the non-actuated and actuated simulations. Figure \ref{fig:cprms} suggests that the actuator excites some of the unstable shear layer modes, triggering shorter wave-length Kelvin-Helmholtz instabilities that (as seen in figure \ref{fig:fft_mic}) would radiate a higher frequency sound when impinging onto the trailing edge of the cavity. Additionally, the directivity of the sound radiation has been altered by the flow actuation. The $p_{rms}$ contours in the controlled case have lower levels in the area surrounding the sensor. Also, the contours show higher values towards the leading edge of the cavity, which indicates an upstream noise propagation.

\section{Conclusions}
\label{sec:6}

A newly developed framework oriented to adjoint-based optimal flow control was successfully applied on a two-dimensional open cavity flow. The target of the actuation was to minimise the pressure fluctuations at the sensor location, and an amplitude reduction of $79.6\%$ respect to the non-actuated case was achieved. A statistical study of the adjoint-field prior to the optimisation indicated the ideal location for the actuator. To exploit the advantages of the adjoint method, a large number of control parameters was used. Decreasing the flow-field sampling frequency acted as an implicit low-pass filter in the sensor. This resulted into an increase in the energy content for frequencies higher than the sampling frequency of the sensor. Finally, $p_{rms}$ contours revealed a decrease of pressure fluctuations in the area surrounding the sensor, by changing the directivity of the noise radiation of the cavity's trailing edge.


\begin{acknowledgements}
The authors are grateful to EPSRC for the funding provided, and UK turbulence consortium under EPSRC grant EP/L000261/1 for the computational resources to carry out the simulations on ARCHER (national UK supercomputing facility). For the purposes of EPSRC data policy, no significant data were generated in this study.
\end{acknowledgements}


\appendix

\section{Adjoint Equations in Characteristic Form}
\label{sec:App_a}

The method of characteristics can be applied to the adjoint equations in the same manner as for the Navier-Stokes equations \cite{thompson1}. Note that eqs. (\ref{eq:adjointEuler1}), (\ref{eq:adjointEuler2}) and (\ref{eq:adjointEuler3}) lack conservative or flux form, hence the procedure gets easier since the conversion to primitive variables is already done \cite{thompson2}. The code used in the present investigation only allows periodic boundary conditions in the spanwise direction, which means that the characteristic equations only need to be derived in the streamwise ($\xi$) and vertical ($\eta$) directions. This appendix only illustrates the derivation of the characteristic equations in the $\xi$ direction on a uniform Cartesian grid. The derivation for the $\eta$ direction proceeds similarly. The idea behind the method is to transform the derivatives that are normal to the computational boundary into combinations of the amplitudes of the characteristic waves $\mathscr{L}$, where each one of them has its own characteristic velocity $\lambda$. The $\mathscr{L}$ values, which are initially computed from the data within the domain, are corrected afterwards by imposing the appropriate conditions. To achieve this transformation, the governing equations first need to be written in the form\begin{equation}
-\frac{\partial Q^*}{\partial t} + A_\xi \frac{\partial Q^*}{\partial \xi} +C_\xi=0,
\label{ceq1}
\end{equation}
where $C_\xi$ contains the non-normal derivatives to the boundary, the source terms and the viscous terms. If the full viscous equations are considered, the system of equations becomes non-hyperbolic, which makes necessary the use of certain assumptions. The LODI relations \cite{poinsotlele} assume that the system is one-dimensional and inviscid at the boundaries, which implies that the contribution of $C_\xi$ can be neglected. Following this strategy, the boundary conditions are `designed' for the LODI problem, but the solution is advanced in time without any assumptions. The matrix $A_\xi$ contains all the coefficients multiplying the normal derivatives to the computational boundary. Hence, the matrix $A_\xi$ can be written as
\begin{equation}
A_\xi = 
\begin{pmatrix}
-u & -1 & 0 & 0 & 0 \\ 
-c^2 & -2u & -v & -w & -1\\
0 & 0 & - u & 0 & 0 \\
0 & 0 & 0 & - u & 0 \\
c^2u & u^2 & uv & uw & 0 \\
\end{pmatrix},
\end{equation}
where the speed of sound $c$ is \begin{equation}
c^2=\frac{\gamma p}{\rho}.
\end{equation}
The characteristic propagation velocities can be obtained by calculating the eigenvalues of $A_\xi$, giving
\begin{equation}
\lambda = 
\begin{pmatrix}
-u \\ -u\\ -u\\ -u-c\\ -u+c\\
\end{pmatrix}.
\end{equation}
At this point it can be seen already that the propagation of the adjoint characteristics occurs (as expected) in the opposite direction to the forward system \cite{kimlee}. The amplitude of the characteristics is defined as \begin{equation}
\mathscr{L} \equiv \lambda \left[ S_\xi^{-1} \frac{\partial Q^*}{\partial \xi}\right]^{\text{T}},
\label{lxi}
\end{equation} where the rows of the matrix $S_\xi^{-1}$ are left eigenvectors of $A_\xi$. The matrix $S_\xi^{-1}$ can be written as
\begin{equation}
S_\xi^{-1} = 
\begin{pmatrix}
0 & u & 0 & 0 & 1\\
0 & 0 & 0 & 1 & 0\\
0 & 0 & 1 & 0 & 0\\
-\frac{c}{2} & - \frac{c + u}{2c} & -\frac{v}{2c} & -\frac{w}{2c} & -\frac{1}{2c}\\
\frac{c}{2} & \frac{-c+u}{2c} & \frac{v}{2c} & \frac{w}{2c} & \frac{1}{2c}\\
\end{pmatrix}.
\end{equation}
At this stage, boundary conditions can be applied by modifiyng $\mathscr{L}$, giving corrected  $\mathscr{L}'$ values. Hence, (\ref{ceq1}) must be rewritten to get the corrected time derivative as \begin{equation}
\label{chareq_xi}
-\frac{\partial Q^*}{\partial t} + S_\xi \mathscr{L}' +C_\xi=0,
\end{equation}
with $S_\xi$ being a matrix whose columns are the right eigenvalues of $A_\xi$. The $S_\xi \mathscr{L}'$ term can be written as
\begin{equation}
S_\xi \mathscr{L}'=
\begin{pmatrix}
-\frac{1}{c^2} & -\frac{w}{c^2} & -\frac{v}{c^2} & -\frac{1}{c} & \frac{1}{c} \\
0 & 0 & 0 & -1 & -1 \\
0 & 0 & 1 & 0 & 0 \\
0 & 1 & 0 & 0 & 0\\
1 & 0 & 0 & u & u\\ 
\end{pmatrix}
\cdot
\begin{pmatrix}
\mathscr{L}'_1\\
\mathscr{L}'_2\\
\mathscr{L}'_3\\
\mathscr{L}'_4\\
\mathscr{L}'_5
\end{pmatrix}.
\end{equation}

\bibliographystyle{spmpsci}      
\bibliography{References.bib}


\end{document}